\documentclass[reprint, amsmath,amssymb, prb,superscriptaddress,longbibliography]{revtex4-2}

\usepackage{graphicx}
\usepackage{dcolumn}
\usepackage{bm}
\usepackage[usenames,dvipsnames]{xcolor}
\usepackage{hyperref}
\usepackage[inline]{enumitem}
\usepackage{bbold}
\usepackage{soul}
\usepackage{placeins}
\usepackage{amsmath}
\usepackage{amssymb}
\usepackage{physics}
\usepackage{placeins}

\newcommand{\up}{\uparrow}

\newcommand{\dn}{\downarrow}

\begin{document}

\title{Properties and prevalence of false poor man’s Majoranas in\\ two- and three-site artificial Kitaev chains}

\author{Melina Luethi}
\affiliation{Department of Physics, University of Basel, Klingelbergstrasse 82, CH-4056 Basel, Switzerland}
\author{Henry F. Legg} 
\affiliation{Department of Physics, University of Basel, Klingelbergstrasse 82, CH-4056 Basel, Switzerland}
\affiliation{SUPA, School of Physics and Astronomy, University of St Andrews,
North Haugh, St Andrews, KY16 9SS, United Kingdom}
\author{Daniel Loss}
\affiliation{Department of Physics, University of Basel, Klingelbergstrasse 82, CH-4056 Basel, Switzerland}
\author{Jelena Klinovaja}
\affiliation{Department of Physics, University of Basel, Klingelbergstrasse 82, CH-4056 Basel, Switzerland}

\date{\today}

\begin{abstract}
It was predicted that a minimal chain of two quantum dots (QDs) connected via a superconductor can host perfectly localized zero-energy states, known as poor man's Majoranas (PMMs). It is expected that these states are related to Majorana bound states (MBSs) in longer chains and that the tunable nature of this setup makes it a promising platform to study MBSs. However, realistic systems can only host highly, but not perfectly, localized near-zero-energy states, called imperfect PMMs. It has been shown that these imperfect PMMs can evolve into trivial states unrelated to MBSs when the chain is extended. Such states are called false PMMs, whereas PMMs that evolve into MBSs in long chains are called true PMMs. Here, using a microscopic model of QD-superconductor arrays, we consider properties of false PMMs and the circumstances under which they appear. In two-site systems, we find that the origin of many false PMMs can be related to zero-energy states occurring in the absence of superconductivity and we use this analytic understanding to characterize the false PMMs that are typical for different regions of parameter space.  In three-site systems, we show that false PMMs can occur via the same mechanism as for two-site systems, but we also find them in regions of parameter space where they are not predicted to exist, thus hinting that the physics of false PMMs can be richer in longer chains. Finally, we demonstrate that the PMMs most stable to perturbations in chemical potential and with the largest excitation gaps appear in a region of parameter space that also has a large ratio of false to true PMMs.
\end{abstract}

\maketitle
\bibliographystyle{apsrev4-1}

\section{\label{sec:intro}Introduction}
Majorana bound states (MBSs) are quasiparticles that emerge in topological superconductors~\cite{kitaev2001unpaired}. It has been proposed that they can store and manipulate quantum information in a fault-tolerant way~\cite{kitaev2003fault, nayak2008non, elliot2015colloquium} due to their non-Abelian exchange statistics~\cite{ivanov2001non}. The Kitaev chain~\cite{kitaev2001unpaired} is a spinless minimal model that hosts MBSs. It relies on $p$-wave superconductivity, however there has not yet been any strong evidence for intrinsic $p$-wave superconductivity and this has precluded a direct implementation. Nanowires with strong spin-orbit interaction (SOI) proximitized by a superconductor~\cite{lutchyn2010majorana, oreg2010helical, stanescu2011majorana, mourik2012signatures, klinovaja2012composite, deng2012anomalous, das2012zero, laubscher2021majorana}  
are prominent examples of an effective implementation of the Kitaev chain. However, despite great experimental effort, there has not yet been a conclusive observation of MBSs in nanowires, largely due to the fact that disorder can result in signals that mimic MBSs~\cite{kells2012near, lee2012zero, rainis2013towards, roy2013topologically, ptok2017controlling, liu2017andreev, moore2018two, moore2018quantized, reeg2018zero, vuik2019reproducing, stanescu2019robust, woods2019zero, chen2019ubiquitous, awoga2019supercurrent, prada2020andreev, yu2021non, sarma2021disorder, valentini2021nontopological, hess2021local,hess2022trivial, hess2022prevalence}. 

\begin{figure}
\centering
\includegraphics[width=\linewidth]{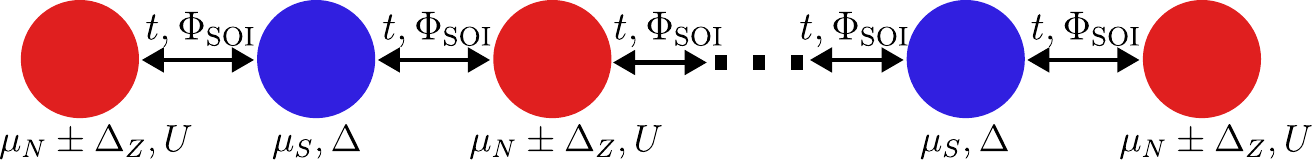}
\caption{ Sketch of a chain of QDs that can host PMMs. The red (blue) QDs are normal (superconducting) QDs. The normal QDs are characterized by a chemical potential $\mu_N$, Zeeman energy $\Delta_Z$, and Coulomb repulsion $U$. The superconducting QDs are characterized by a chemical potential $\mu_S$ and a superconducting pairing $\Delta$. 	The hopping between the QDs is characterized by the amplitude $t$ and the SOI angle $\Phi_\mathrm{SOI}$. 	The minimal chain consists of two normal QDs and one superconducting QD. When going to longer chains, we assume a uniform chain and that there is no superconducting phase difference between the QDs.}
\label{fig:setup}
\end{figure}

To overcome the issue of disorder, implementing the Kitaev chain as an array of quantum dots (QDs) has been proposed~\cite{sau2012realizing, leijnse2012parity, fulga2013adaptive, su2017andreev, tsintzis2022creating, liu2022tunable, dvir2023realization, sanches2023fractionalization, chunxiao2023fusion, koch2023adversarial, souto2023probing, zatelli2023robust, samuelson2023minimal, tsintzis2023roadmap, pino2024minimal, luna2024fluxtunable, bordin2024signatures, boross2023braiding, liu2024coupling, vilkelis2024fermionic, bozkurt2024interaction, vandriel2024crossplatform, geier2023fermionparity, haaf2023engineering, kocsis2024strong, cayao2024emergent, miles2024kitaev, cayao2024nonhermitian, liu2024protocol, alvarado2024interplay, svensson2024quantum, gomez2024high, pan2024rabi, pandey2024dynamics, liu2023enhancing, benestad2024machine, ezawa2024even, luethi2024perfect, luethi2024fate, kotetes2024nonreciprocal, ten2024edge, huang2024unified, souto2024majorana, nitsch2024poor}, see Fig.~\ref{fig:setup}. 
The QDs are separated by superconducting sections that transmit elastic cotunneling (ECT) and crossed Andreev reflection (CAR) between pairs of QDs~\cite{recher2001andreev,lesovik2001electronic,falci2001correlated, bouchiat2002single, feinberg2003andreev, hofstetter2009cooper, herrmann2010carbon}. It has been predicted that, by tuning the relative strengths of CAR and ECT, the chain can be brought to a ``sweet spot'', i.e., a point in parameter space where a pair of states, with its partners localized at opposite ends of the chain, do not overlap and have the properties of Majorana bound states, even in a minimal chain consisting only of two QDs~\cite{sau2012realizing, leijnse2012parity}. In recent years, control over the strengths of ECT and CAR has been demonstrated in multiple experiments~\cite{liu2022tunable, wang2022singlet, kurtossy2022parallel, dejong2023controllable, bordin2023tunable, wang2023triplet, bordin2024crossed}. In minimal chains, due to the lack of topological protection, the states appearing at these sweet spots have been called ``poor man's Majoranas'' (PMMs). However, these sweet spots only exist in highly simplified models, where it is commonly assumed that the QDs are fully spin-polarized~\cite{sau2012realizing, leijnse2012parity} and/or there is a high degree of independence between parameters~\cite{luethi2024perfect}. In more realistic models highly -- but not perfectly -- localized near-zero-energy states exist~\cite{tsintzis2022creating, luethi2024perfect}, which have been referred to as ``imperfect PMMs''~\cite{luethi2024perfect}. 
It has been shown that when the array of QDs is extended uniformly to the long chain limit, imperfect PMMs can evolve either into true MBSs, or into trivial states~\cite{luethi2024fate}. The former are called ``true PMMs'', whereas the latter ``false PMMs''~\cite{luethi2024fate}. It is important to emphasize that the extension of the chain using uniform parameters is a theoretical tool and in realistic experimental systems parameters will vary along the chain, which will further complicate whether a PMM evolves into a topological state in the long chain limit.

While the existence of false PMMs was demonstrated in Ref.~\cite{luethi2024fate}, we study here the properties of false PMMs and ascertain the regions of parameter space where they occur. To do this, we utilize a microscopic model for the chain shown in Fig.~\ref{fig:setup}. First, studying the system in the absence of superconductivity and putting constraints on the energy as well as the wave functions, we determine analytically where in parameter space false PMMs appear in two-site chains. We find that there are two defining characteristics of false PMMs in two-site chains: 1) Whether they appear at Zeeman energies smaller or larger than the range of Zeeman energies within which the uniform long chain hosts a topological phase and 2) whether the false PMM is associated with a trivial zero-energy crossing (ZEC) in the long chain limit. Throughout, we compare our analytical results to numerical calculations and find  good agreement. 

We also consider three-site chains, which have been studied in recent experiments~\cite{bordin2024signatures, ten2024edge}, and show that the same mechanism as in the two-site chains can cause false PMMs in these systems. We again compare our analytical results to numerical calculations and find  good agreement. 
In addition, we also  find false PMMs that cannot be explained by the same mechanism as the two-site system.
This emphasizes that further mechanisms for false PMMs can be present as the chain is extended.

Finally, we study the ratio of true to false PMMs as a function of the threshold values that are required to define an imperfect PMM~\cite{luethi2024perfect}. As would be expected, we find that stricter threshold values result in fewer states overall being classified as PMMs, but also that the ratio of true to false PMMs increases as the threshold values are made more restrictive. Additionally, we find that PMMs most stable to perturbations in chemical potential and with the largest excitation gaps occur at large inter-QD hopping amplitudes. However, for these large hopping amplitudes, the PMMs are not as well localized and therefore threshold values cannot be too stringent if these states should satisfy the PMM condition. As a result, in this regime, the ratio of false PMMs is rather high.

The rest of this paper is structured as follows. In Sec.~\ref{sec:model}, we introduce the microscopic model of an artificial Kitaev chain and the quantities required to characterize imperfect PMMs. Next,  in Sec.~\ref{sec:classification}, we study the properties of false PMMs in minimal and longer QD chains in different regions of parameter space. In Sec.~\ref{sec:prevalence_2_sites} we analyze how the ratio of true and false PMMs in minimal QD chains behaves as a function of the threshold values and study the stability of the PMMs throughout parameter space. We conclude in Sec.~\ref{sec:conclusion}. 
In Appendix~\ref{app:large_zeeman}, we address the finite Zeeman energy in the superconducting section. The effect of finite on-site Coulomb repulsion is studied in Appendix~\ref{app:with_U}.
We compare analytically and numerically calculated optimal chemical potentials in Appendix~\ref{app:analytical_vs_optimized_mu}.

\section{\label{sec:model}Model}
To model an artificial Kitaev chain, we consider a chain of alternating normal and superconducting QDs, see Fig.~\ref{fig:setup}, as was first introduced in Ref.~\citenum{tsintzis2022creating}. 
The normal QDs have chemical potential $\mu_N$, Zeeman energy $\Delta_Z$, and on-site Coulomb repulsion $U$. The superconducting QDs model Andreev bound states (ABSs) that effectively transmit ECT and CAR between the normal QDs. The superconducting QDs have a superconducting pairing potential $\Delta$ and chemical potential $\mu_S$. Both $\mu_N$ and $\mu_S$ are measured with respect to the chemical potential of the superconductor hosting the ABS.
Due to screening by the parent superconductor, we assume that the superconducting QDs have no Zeeman energy and no Coulomb repulsion. It was shown in Ref.~\citenum{tsintzis2022creating} that this assumption is not necessary for PMMs and we comment on the effect of finite Zeeman energy on the superconducting QDs in Appendix~\ref{app:large_zeeman}. 
We assume that all parameters are uniform throughout the chain and that there is no superconducting phase difference between the superconducting QDs. 
In the following, we will distinguish between a minimal chain that consists of two normal (and one superconducting) QDs, chains that consist of three normal (and two superconducting) QDs, and long chains that consist of several tens to hundreds of QDs.

Let us now consider a chain with $N$ normal QDs and $N-1$ superconducting QDs. We label each QD with an index $j = 0, 1, \dots , 2N-2$ and if $j$ is even (odd), then the corresponding QD is normal (superconducting).
The hopping between the QDs is determined by the hopping amplitude $t$ and the SOI angle $\Phi_\mathrm{SOI}$. 
The chain is aligned along the $x$ axis and we apply a magnetic field parallel to the $z$ axis, leading to a Zeeman energy $\Delta_Z$. Thus, the Hamiltonian of the system is written as
\begin{align}  
	& H = \sum_{j=0}^{2N-2}  \Big[ 
	\sum_{\sigma=\up,\dn} 
	(\mu_j +  \sigma \Delta_{Z,j}  ) n_{j\sigma}
	 +  U_j n_{j\up} n_{j \dn}  
	\nonumber \\ & 
	+ \Delta_j (d_{j\up}^\dagger d_{j\dn}^\dagger + d_{j\dn} d_{j\up} )	
	\Big]
	\nonumber \\ & 
	 + t \!\! \sum_{j=0}^{2N-3}  \sum_{\sigma,\sigma'=\up,\dn} \!\! \Big[ 
	 U_\mathrm{SOI} \! \left(\frac{\Phi_\mathrm{SOI}}{2}\right)_{\!\!\! \sigma\sigma'} 
	 d_{j+1 \sigma}^\dagger d_{j \sigma'} 
	+ \mathrm{H.c.} \Big] \!,
	\label{eq:full_hamiltonian}
\end{align}
where $n_{j \sigma}=d_{j \sigma}^\dagger d_{j \sigma}$, $d_{j \sigma}^\dagger$ ($d_{j \sigma}$) creates (annihilates) a particle on QD $j$, the notation $\sigma \Delta_Z$ means $+\Delta_Z$ for $\sigma=\up$ and $-\Delta_Z$ for $\sigma=\dn$, $U_j$ is the on-site repulsion strength, and
\begin{subequations}
	{\allowdisplaybreaks
\begin{align} \allowdisplaybreaks
\mu_j =& \begin{cases}
\mu_N & \mathrm{if} \, j \, \text{even}, \\
\mu_S & \mathrm{if} \, j \, \text{odd}, 
\end{cases} \\
\Delta_{Z,j} =& \begin{cases}
	\Delta_Z & \mathrm{if} \, j \, \text{even}, \\
	0 & \mathrm{if} \, j \, \text{odd}, 
\end{cases} \label{eq:definition_delta_z} \\
U_j =& \begin{cases}
U & \mathrm{if} \, j \, \text{even}, \\
0 & \mathrm{if} \, j \, \text{odd} ,
\end{cases} \\
\Delta_j =& \begin{cases}
0 & \mathrm{if} \, j \, \text{even}, \\
\Delta & \mathrm{if} \, j \, \text{odd} .
\end{cases} 
\end{align} 
}
\end{subequations}
The SOI matrix is given by~\cite{spethmann2023highfidelity, luethi2024perfect}
\begin{align}
	U_\mathrm{SOI} \left(\frac{\Phi_\mathrm{SOI}}{2}\right) = \cos \left(\frac{\Phi_\mathrm{SOI}}{2}\right) + i \sin \left(\frac{\Phi_\mathrm{SOI}}{2}\right) \sigma_y,
\end{align}
where $\sigma_y$ is the second Pauli matrix and we assume $0 \leq \Phi_\mathrm{SOI} \leq \frac{\pi}{2}$~\footnote{
The assumption $0 \leq \Phi_\mathrm{SOI} \leq \frac{\pi}{2}$ is justified because any effective coupling between two normal QDs is the result of at least two consecutive hoppings, i.e., SOI will appear as the term $U_\mathrm{SOI} (\Phi_\mathrm{SOI})$. Furthermore, $U_\mathrm{SOI} (\Phi_\mathrm{SOI}) = - U_\mathrm{SOI} (\pi - \Phi_\mathrm{SOI})^T = -U_\mathrm{SOI} (\pi + \Phi_\mathrm{SOI}) $.  Therefore,  it is sufficient to study SOI angles in the interval $ 0 \leq \Phi_\mathrm{SOI} \leq \frac{\pi}{2}$.}.
For the main part of this work, we set the on-site repulsion such that $U_j=0$ for all $j$, including the normal QDs. We consider the case of $U>0$ in Appendix~\ref{app:with_U}. However, in the case $U=0$,  the energy spectrum of the chain can also be calculated in the BdG formulation of Eq.~\eqref{eq:full_hamiltonian} and we will do this when calculating the spectrum in the long chain limit. Furthermore, if $U_j=0$ for all $j$, then the BdG Hamiltonian can be extended to describe an infinitely long chain in momentum space and the topological invariant $W$ can be calculated, see Ref.~\citenum{luethi2024fate}. A state is trivial (topological) if $W=1$ ($W=-1$). 

For all other calculations 
we use the second quantized form of Eq.~\eqref{eq:full_hamiltonian}, for which we define the basis $\mathcal{B} = \{\ket{n_{0 \up}, n_{0 \dn}, n_{1 \up}, n_{1 \dn}, \dots, n_{2N-2 \up}, n_{2N-2 \dn}} : n_{j\sigma}=0,1\}$. In this basis, the Hamiltonian is described by the matrix $\bra{\psi} H \ket{\chi}$, where $\ket{\psi}, \ket{\chi} \in \mathcal{B}$, and $H$ is the Hamiltonian defined in Eq.~\eqref{eq:full_hamiltonian}.
Since the Hamiltonian conserves the particle number parity, it splits into a block-diagonal matrix with sub-matrices $\mathcal{H}_\mathrm{even}$ and $\mathcal{H}_\mathrm{odd}$, with even and odd particle number parity, respectively.
We label the eigenvectors of $\mathcal{H}_\mathrm{even}$ ($\mathcal{H}_\mathrm{odd}$) as $\ket{\Psi_a^\mathrm{even}}$ ($\ket{\Psi_a^\mathrm{odd}}$) and their corresponding eigenvalues as $E_a^\mathrm{even}$ ($E_a^\mathrm{odd}$), where $a$ numbers the eigenvalues, ordered such that $E_0^\mathrm{even} \leq E_1^\mathrm{even} \leq E_2^\mathrm{even} \leq \dots$ ($E_0^\mathrm{odd} \leq E_1^\mathrm{odd} \leq E_2^\mathrm{odd} \leq \dots$). 
We introduce the energy difference~\cite{leijnse2012parity}
\begin{equation} \label{eq:definition_dE} 
	\Delta E = E_0^\mathrm{even} - E_0^\mathrm{odd},
\end{equation}
the charge difference on QD $j$~\cite{tsintzis2022creating}
\begin{equation} \label{eq:definition_dQ}
	\!\!\!\!\! \Delta Q_j \!=\! \sum_{\sigma} \! \big( \!
	\bra{\Psi_0^\mathrm{even}} \! n_{j\sigma} \! \ket{\Psi_0^\mathrm{even}}
	\!-\! \bra{\Psi_0^\mathrm{odd}} \! n_{j\sigma} \! \ket{\Psi_0^\mathrm{odd}}
	\!
	\big),
\end{equation}
the Majorana polarization (MP) on QD $j$~\cite{aksenov2020strong, tsintzis2022creating, samuelson2023minimal},
\begin{align} \label{eq:definition_M} 
	&M_j = \frac{
		\left|\sum_{\sigma} \sum_{s=\pm 1} 
		\bra{\Psi_0^\mathrm{even}} \eta_{j \sigma s} \ket{\Psi_0^\mathrm{odd}}^2
		\right|
	}{
		\sum_{\sigma} \sum_{s=\pm 1}  \left|
		\bra{\Psi_0^\mathrm{even}} \eta_{j \sigma s} \ket{\Psi_0^\mathrm{odd}}^2
		\right|
	}, \nonumber \\
	&\eta_{j\sigma +} = d_{j\sigma} + d_{j\sigma}^\dagger, \,
	\eta_{j\sigma -} = i \left( d_{j\sigma} - d_{j\sigma}^\dagger \right),
\end{align}
and the excitation gap
\begin{equation} \label{eq:definition_gap} 
	E_\mathrm{ex} = \min \{
	E_1^\mathrm{even} - E_0^\mathrm{even},
	E_1^\mathrm{odd} - E_0^\mathrm{odd}
	\}.
\end{equation}
Here, introducing the MP, we used that the fact that our Hamiltonian $H$ is real and, thus, the corresponding wave functions are real as well.
Since we are mostly only interested in the charge difference and MP on the first ($j=0$) and last ($j=2N-2$) QDs, we define $\Delta Q_L = \Delta Q_0$, $\Delta Q_R = \Delta Q_{2N-2}$, $M_L = M_0$, and $M_R = M_{2N-2}$.
Originally, PMMs were introduced in Refs.~\cite{sau2012realizing, leijnse2012parity} as states in minimal chains, i.e., $N=2$, with $\Delta E = 0$, $\Delta Q_L = \Delta Q_R = 0$, $M_L=M_R=1$, and $E_\mathrm{ex}>0$. We call such states perfect PMMs~\cite{luethi2024perfect}. However, as shown in Ref.~\cite{luethi2024perfect}, perfect PMMs cannot exist in the model discussed here. Instead, we define a threshold region (TR), in which so-called imperfect PMMs, i.e., highly localized near-zero energy states, exist~\cite{luethi2024perfect}:
\begin{align} \label{eq:definition_ROT}
	&|\Delta E | < \Delta E_\mathrm{th}
	\text{ and }
	E_\mathrm{ex} > E_\mathrm{ex, th} 
	\nonumber \\ &
	\text{ and }
	|\Delta Q_L | < \Delta Q_\mathrm{th}
	\text{ and }
	|\Delta Q_R | < \Delta Q_\mathrm{th}
	\nonumber \\ &
	\text{ and }
	M_L > 1-M_\mathrm{th}
	\text{ and }
	M_R > 1-M_\mathrm{th},
\end{align}
where $\Delta E_\mathrm{th}$, $E_\mathrm{ex, th} $, $\Delta Q_\mathrm{th}$, and $M_\mathrm{th}$ are threshold values to be chosen.  
Expanding to the long chain limit, some imperfect PMMs evolve into topologically protected MBSs, which we call true PMMs. On the other hand, imperfect PMMs may evolve into trivial states~\cite{luethi2024fate} and we call these false PMMs. Furthermore, in Ref.~\citenum{luethi2024fate}, the term ``scaled PMMs'' was introduced for highly localized near-zero-energy states in short chains with $N>2$ sites. In this work, we will not make this distinction explicit and simply refer to PMMs in both cases.

\section{\label{sec:classification}Analytical study of false PMMs}
\begin{figure}
	\centering
	\includegraphics[width=\linewidth]{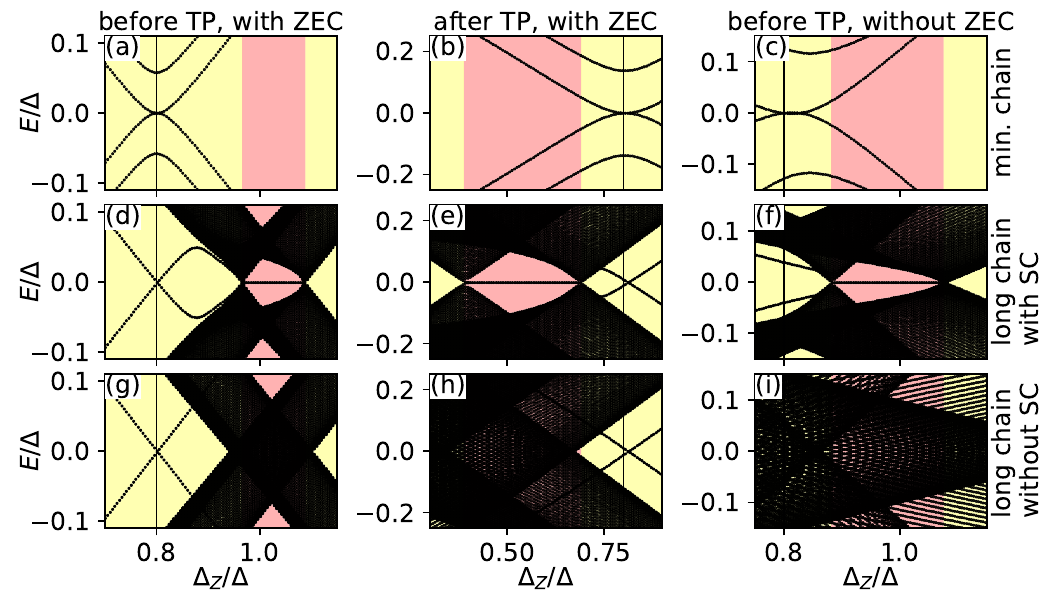}
	\caption{
		Examples of the three categories of false PMMs. (a)-(c): Energy spectrum in the minimal chain. All examples have a highly localized near-zero-energy state at $\Delta_Z/\Delta = 0.8$ that satisfies the TR condition of Eq.~\eqref{eq:definition_ROT}. The background color indicates the topological invariant in the infinite chain: a yellow (red) background indicates that the infinite chain is trivial (topological). All three PMMs at $\Delta_Z/\Delta = 0.8$ lie in the trivial regime, thus these are false PMMs.
		(d)-(f): Energy spectrum in a long uniform chain ($N=100$) with the same parameters as in  panels (a)-(b), respectively. The false PMMs in panels (d) and (e) are associated with a fourfold degenerate ZEC, the false PMM in panel (f) is not. The false PMMs in panels (d) and (f) [panel (e)] are at smaller [larger] Zeeman energies than what is required for the long chain to enter the topological phase. 
		(g)-(i): Energy spectrum in the same long uniform chain as for (d)-(f), but now without any superconductivity. The ZEC that were present in panels (d) and (e) are still present in panels (g) and (h), even without superconductivity.
		The parameters are (rounded to three significant digits) as follows. (a), (d), (g):
		$t/\Delta = 1.02$, $\Phi_\mathrm{SOI}/\pi = 0.452$, $\mu_N/\Delta=0.624$, $\mu_S/\Delta=-5.06$.
		(b), (e), (h): $t/\Delta=1.49$, $\Phi_\mathrm{SOI}/\pi = 0.425$, $\mu_N/\Delta=1.21$, $\mu_S/\Delta=6.35$.
		(c), (f), (i): $t/\Delta=0.637$, $\Phi_\mathrm{SOI}/\Delta=0.0706$, $\mu_N/\Delta=0.891$, $\mu_S/\Delta=0.706$.
		All three false PMMs have $|\Delta E| < 10^{-8}$, $E_\mathrm{ex}>0.05$, $|\Delta Q_{L}|=|\Delta Q_R| < 10^{-7}$, and $1-M_L = 1-M_R < 0.11$.
	}
	\label{fig:false_pmms_examples}
\end{figure}

As mentioned above, we observe two characteristics of false PMMs, based on their spectrum $E(\Delta_Z)$ in the long chain limit (see Fig.~\ref{fig:false_pmms_examples}). 
In this section, we will use both constraints on the energy and wave functions to distinguish the origin and regions of parameter space that host false PMMs with these characteristics. 
The first characteristic is that the false PMMs appear either at Zeeman energies smaller or larger than the range of Zeeman energies required for the uniform long chain in our model to be in the topological phase (TP). The second characteristic is whether the false PMM is associated with a four-fold degenerate zero-energy crossing (ZEC). In principle, these two properties allow for four different categories of false PMMs: ``before TP, with ZEC'', ``after TP, with ZEC'', ``before TP, without ZEC'', ``after TP, without ZEC'', see Fig.~\ref{fig:false_pmms_examples}. However, as we will explain below, we never find any PMM of the category ``after TP, without ZEC''.

As mentioned above, in the following, we set $U=0$ and comment on the case $U>0$ in Appendix~\ref{app:with_U}. We find values for the chemical potentials $\mu_N$ and $\mu_S$ that result in a false PMMs, following two approaches. First, we derive $\mu_{N,\mathrm{a}}$ and $\mu_{S,\mathrm{a}}$ using analytical arguments. Second, we obtain $\mu_{N,\mathrm{n}}$ and $\mu_{S,\mathrm{n}}$ using the numerical optimization algorithm explained in Ref.~\citenum{luethi2024perfect}.
We note that the analytically and numerically obtained chemical potentials are not necessarily equal, but we expect that they are similar and we check this assumption in Appendix~\ref{app:analytical_vs_optimized_mu}.  
We also emphasize that the position of the ZEC in the long chain limit is not necessarily at the Zeeman energy, where the PMM is found in the short chain. Finally, we note that the transition to the TP also results in a closing of the gap in the bulk spectrum, and this can (occasionally) cause an ambiguity when the transition and a potential ZEC are close in Zeeman energy.

\subsection{\label{sec:classification_two_site}False PMMs in two-site chains}
We first consider the minimal chain with $N=2$ (three dots in total). Since $U=0$, as above, the Hamiltonian from Eq.~\eqref{eq:full_hamiltonian} can be written in the BdG form, giving $\mathcal{H}_\mathrm{BdG}$.
To find PMMs, we first search for zero-energy solutions of the minimal chain, thus setting $\det(\mathcal{H}_\mathrm{BdG})=0$, which implies
\begin{align}
	0 \!=  &
	( 
	\mu_{N,\mathrm{a}}^2 \!\! - \! \Delta_Z^2
	\! ) [2t^4
	\! - \! 4 t^2 \mu_{N,\mathrm{a}} \mu_{S,\mathrm{a}} \!+\!  (\mu_{S,\mathrm{a}}^2 \!  + \! \Delta^2)(
	\mu_{N,\mathrm{a}}^2 \! - \! \Delta_Z^2
	)] 	\nonumber \\ &
	+ 2t^4 [\mu_{N,\mathrm{a}}^2- \!  \Delta_Z^2 \cos \left(2 \Phi_\mathrm{SOI}\right)]
	. \label{eq:mu_muM_eq2}
\end{align}
If Eq.~\eqref{eq:mu_muM_eq2} is approximately fulfilled, then the minimal chain hosts a state that is (almost) at zero-energy and so could be an (imperfect) PMM. 
In particular, we can solve Eq.~\eqref{eq:mu_muM_eq2} for $\mu_{S,\mathrm{a}}$. To get a real solution, i.e., $\mu_{S,\mathrm{a}} \in \mathbb{R}$, the following condition must be fulfilled
\begin{align}
4 t^4 \Delta_Z^2 \cos^2 (\Phi_\mathrm{SOI}) \geq \Delta^2 ( \Delta_Z^2 - \mu_{N,\mathrm{a}}^2 )^2 .
\label{eq:condition_muM_real}
\end{align} 
In particular, the left-hand side of Eq.~\eqref{eq:condition_muM_real} vanishes as $\Phi_\mathrm{SOI} \rightarrow \frac{\pi}{2}$ and so, for given $t$, $\Phi_\mathrm{SOI}$, $\Delta_Z$, and $\Delta$ we must have $\mu_\mathrm{N,\mathrm{a}} \approx \Delta_Z$ (where we choose $\mu_\mathrm{N,\mathrm{a}}, \Delta_Z > 0$). 
We note that in this limit, $\mu_{S,\mathrm{a}}$ diverges, such that Eq.~\eqref{eq:mu_muM_eq2} is satisfied. 
Of course, for other values of $\Phi_\mathrm{SOI}$, Eq.~\eqref{eq:mu_muM_eq2} can still be solved. However, the solution is then more involved.

Although Eq.~\eqref{eq:mu_muM_eq2} governs whether there is a \mbox{(near-)zero-}energy state, it is important to note that these states are not necessarily (imperfect) PMMs since we do not yet know much about their localization. In particular, to determine if these states are imperfect PMMs, we also require a constraint on the wave function. In general this is a complicated task that cannot be easily achieved analytically.
However, it was pointed out in Ref.~\citenum{luethi2024fate} that a zero-energy state localized on the first site of a double QD system without superconductivity can be associated with false PMMs. 
Since this state is mainly localized on the first QD, introducing superconductivity on the second QD has only little impact on the state. Furthermore, adding more QDs to the right-hand side of the system has only a small effect on a zero-energy state localized on the leftmost QD.
The Hamiltonian of such a double QD system is
\begin{subequations} \label{eq:h_2qd}
	\begin{align}  
		H_\mathrm{2QD} =&
		(
			d_{1 \up}^\dagger , d_{1 \dn}^\dagger , d_{2 \up}^\dagger , d_{2 \dn}^\dagger
		)
		\mathcal{H}_\mathrm{2QD}
		\begin{pmatrix}
			d_{1 \up} \\ d_{1 \dn} \\ d_{2 \up} \\ d_{2 \dn}
		\end{pmatrix} \\
		\mathcal{H}_\mathrm{2QD} =&
		(\mu_N + \Delta_Z \sigma_z) \frac{\eta_0+\eta_z}{2}
		+ \mu_S \frac{\eta_0-\eta_z}{2} 
		\nonumber \\ &
		+ t \cos \left(\frac{\Phi_\mathrm{SOI}}{2}\right) \eta_1 
		- t \sin \left(\frac{\Phi_\mathrm{SOI}}{2}\right) \sigma_y \eta_y,
	\end{align}
\end{subequations}
where $\sigma_i$ ($\eta_i$) are the Pauli matrices acting in spin (position) space and $\eta_0$ is the unit matrix.
We note that we do not include a superconducting pairing since we are looking for states localized on the first QD which, in our full Hamiltonian, also does not have an SC pairing potential. For this normal double QD system to host a zero-energy state we require $\det (\mathcal{H}_\mathrm{2QD})=0$, which implies 
\begin{equation}
	0 \! = \! \det (\mathcal{H}_\mathrm{2QD}  )
	\! =  
	(t^2-\mu_{N,\mathrm{a}} \mu_{S,\mathrm{a}})^2 - \mu_{S,\mathrm{a}}^2 \Delta_Z^2 
	.\label{eq:mu_muM_eq1} 
\end{equation}
Solving, this gives $\mu_{S,\mathrm{a}} = \frac{t^2}{\mu_{N,\mathrm{a}} \pm \Delta_Z}$ as a constraint.

To find analytically states that will become false PMMs, we look for localized states that satisfy both Eqs.~\eqref{eq:mu_muM_eq2} and~\eqref{eq:mu_muM_eq1} for $\mu_{N,\mathrm{a}}$ and $\mu_{S,\mathrm{a}}$, with given $t$, $\Phi_\mathrm{SOI}$, $\Delta_Z$, and $\Delta$. We emphasize that there are solutions in each quadrant of the $\mu_N$ versus $\mu_S$ parameter space, but we will focus only on the case $\mu_{N,\mathrm{a}}>0$ and consider $\mu_{S,\mathrm{a}}>0$ and $\mu_{S,\mathrm{a}}<0$ separately.

\begin{figure}
	\centering
	\includegraphics[width=\linewidth]{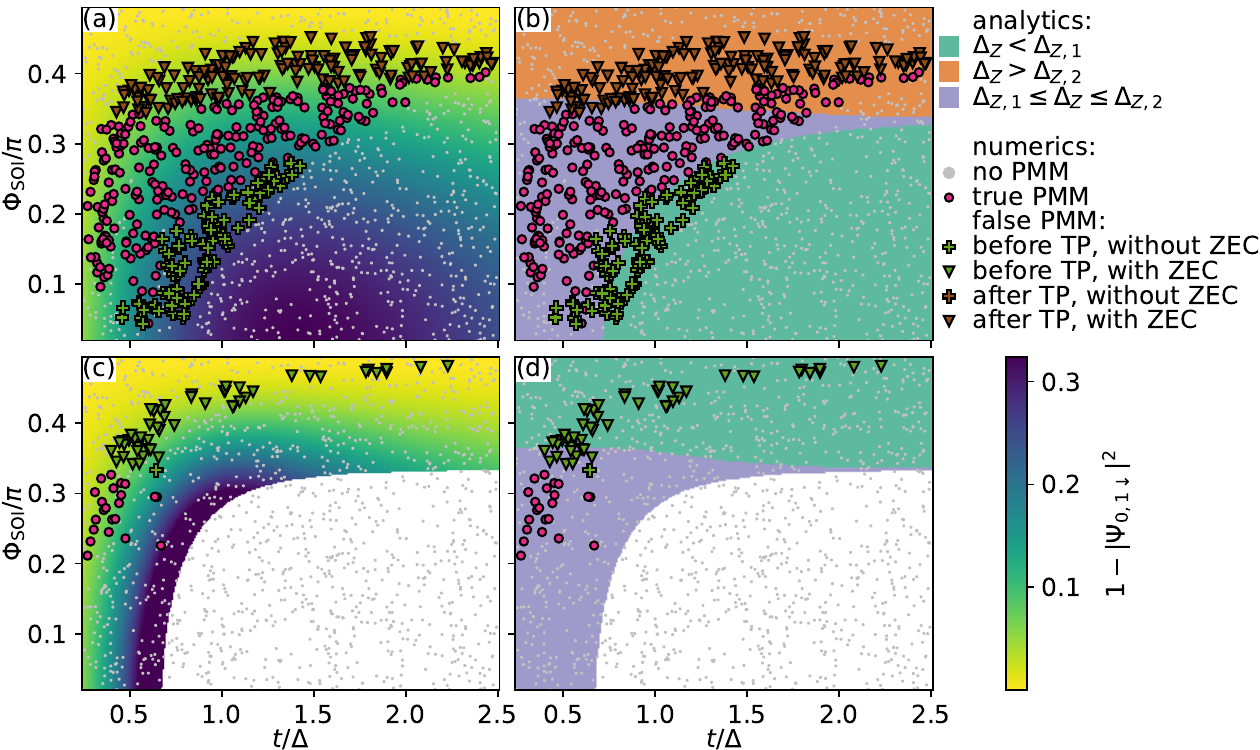}
	\caption{ 
Classification of true (circles) and false PMMs (triangles/crosses) in a minimal chain. The circles, triangles, and crosses 
		in the panels are PMMs classified according to Fig.~\ref{fig:false_pmms_examples} after the chemical potentials $\mu_{N/S,\mathrm{n}}$ are determined using a numerical optimization algorithm~\cite{luethi2024perfect}, requiring $\mu_{N,\mathrm{n}}>0$ and $\mu_{S,\mathrm{n}} > 0$ [$\mu_{S,\mathrm{n}} < 0$] in panels~(a) and~(b) [(c) and~(d)].
Throughout, we keep the Zeeman energy fixed at $\Delta_Z/\Delta = 0.8$. We note that panels~(a) and~(b) [(c) and~(d)] show the same numerical results but different analytical results in the background color.  In panels~(a) and~(c), the background color indicates $|\Psi_{0,1\dn}|^2$ found using Eq.~\eqref{eq:2QD_eigenvector} for the analytically found chemical potentials $\mu_{N/S,\mathrm{a}}$ that solve Eqs.~\eqref{eq:mu_muM_eq2} and~\eqref{eq:mu_muM_eq1}. If the degree of the localization is high, i.e., $|\Psi_{0,1\dn}|^2 \approx 1$, then we expect the PMM to be associated with a ZEC. We see that this agrees with the numerical results, where the triangle symbols associated with a ZEC are located in regions where $|\Psi_{0,1\dn}|^2 \approx 1$.
In panels~(b) and~(d), the background color indicates whether the PMM appears at Zeeman energies smaller or larger than the Zeeman energy required for the chain to enter the topological phase. The Zeeman energies, at which the topological phase transitions  should occur in the infinite chain, are given by $\Delta_{Z,1}<\Delta_{Z,2}$ [see Eq.~\eqref{eq:calculate_zeeman_phase_transition} with analytically found chemical potentials $\mu_{N/S,\mathrm{a}}$ that solve Eqs.~\eqref{eq:mu_muM_eq2} and~\eqref{eq:mu_muM_eq1}] and they are compared to the fixed Zeeman energy $\Delta_Z/\Delta = 0.8$ at which we observe the PMMs.
		If $\Delta_Z < \Delta_{Z,1} < \Delta_{Z,2}$ ($\Delta_{Z,1} < \Delta_{Z,2} < \Delta_Z$), then the PMMs belong to the category ``before TP'' (``after TP''). The corresponding green (orange) background color agrees well with the classification of the numerically optimized and classified PMMs (orange and green symbols).
		If $\Delta_{Z,1} \leq \Delta_Z \leq \Delta_{Z,2}$, then the state is a true PMMs and the corresponding purple background corresponds well with the numerically determined true PMMs (pink symbols).
		In the regions with a white background, no solution for both Eqs.~\eqref{eq:mu_muM_eq2} and~\eqref{eq:mu_muM_eq1} exists.
		The threshold values for the TR are $\Delta E_\mathrm{th}/\Delta = 10^{-4}$, $\Delta Q_\mathrm{th} = 0.01$, $M_\mathrm{th} = 0.3$, and $E_\mathrm{ex,th}/\Delta = 0.05$. The gray dots indicate states that do not satisfy the threshold conditions after the optimization. 
	}
	\label{fig:classification}
\end{figure}

Let us consider the solution $\mu_{S,\mathrm{a}} = \frac{t^2}{\mu_{N,\mathrm{a}} - \Delta_Z}$ of Eq.~\eqref{eq:mu_muM_eq1}. The corresponding zero-energy eigenvector in the double QD has the form 
{\allowdisplaybreaks
\begin{align} 
	\!\! \Psi_0 \!\! =\!\!
	\frac{1}{\mathcal{N}}
	\Big( \!
		0, t, \sin \!\! \left( \!\! \frac{\Phi_\mathrm{SOI}}{2} \!\! \right) \!\! (\mu_{N,\mathrm{a}}  \!-\!  \Delta_Z ) , 
		\cos \!\! \left( \!\! \frac{\Phi_\mathrm{SOI}}{2} \!\! \right) \!\! (\Delta_Z   \!-\! \mu_{N,\mathrm{a}} )
	\!\Big)^T \!\!\!,  
\label{eq:2QD_eigenvector}
\end{align}
with 
$\mathcal{N} = \! \sqrt{
	t^2 + (\mu_{N,\mathrm{a}}-\Delta_Z)^2} $.
}
We see from Eq.~\eqref{eq:2QD_eigenvector} that, if $\mu_{N,\mathrm{a}} \approx \Delta_Z$, then $\Psi_{0,1\dn} = (0,1,0,0) \cdot \Psi_0$ becomes the dominant term, i.e.,  
$|\Psi_{0,1\dn}|^2 \approx 1$, and therefore the state is mainly localized on the first QD. Since the state is both well localized and at zero energy for two QDs, adding further sites and superconductivity to achieve the setup of Fig.~\ref{fig:setup} has little impact on its energy and wave function. We also saw above that in the case $\Phi_{\rm SOI}\approx \frac{\pi}{2}$ the solution $\mu_{N,\mathrm{a}} \approx \Delta_Z$ can readily satisfy the requirements for a (near-)zero-energy state in the minimal chain. In other words, since the state is well localized and close to zero-energy, we can expect a false PMM in the minimal chain. Furthermore, we expect that this PMM is related to a ZEC in the long chain limit. Indeed, we find that regions of parameter space where $|\Psi_{0,1\dn}|^2 \approx 1$ host false PMMs associated with a ZEC, see Fig.~\ref{fig:classification}. 
It is important to emphasize that this ZEC occurs even without the presence of superconductivity, see Fig.~\ref{fig:false_pmms_examples}(g) and (h), and therefore it is unrelated to topological superconductivity.

Numerically [see Fig.~\ref{fig:classification}], we also find false PMMs in regions of parameter space in which the analytical solution of Eqs.~\eqref{eq:mu_muM_eq2} and~\eqref{eq:mu_muM_eq1} is away from the limit $|\Psi_{0,1\dn}|^2 \approx 1$ discussed above. In these regions of parameter space, we find false PMMs without a ZEC. The reason for this is that once we deviate from the limit $|\Psi_{0,1\dn}|^2 \approx 1$, the assumption that PMMs satisfy both Eqs.~\eqref{eq:mu_muM_eq2} and~\eqref{eq:mu_muM_eq1} is not justified anymore. Instead, the numerical optimization algorithm introduced in Ref.~\cite{luethi2024perfect} searches a solution to Eq.~\eqref{eq:mu_muM_eq2} with a maximized MP. These parameters will give much less localized states with finite energy in the double QD system defined in Eq.~\eqref{eq:h_2qd}. Therefore, we do not necessarily expect the numerically optimized chemical potentials $\mu_{N/S,\mathrm{n}}$ to agree with the analytically calculated chemical potentials $\mu_{N/S,\mathrm{a}}$. Nevertheless, we demonstrate in Appendix ~\ref{app:analytical_vs_optimized_mu} that the analytically calculated chemical potentials and the numerically optimized chemical potentials are still relatively close to each other, which justified the usage of analytical values in Fig. \ref{fig:classification} over the entire parameter space.

The second characteristic used to classify false PMMs is whether they appear for Zeeman energies larger or smaller than the Zeeman energy that is required to bring the long chain into the topological phase. 
The long chain goes through two topological phase transitions, one is characterized by a closing of the bulk energy gap at $k=0$, the other one by a bulk energy gap closing at $k=\pi$. 
We calculate the bulk energy spectrum with the momentum-space Hamiltonian $\mathcal{H}(k)$, which is given in the SM of Ref.~\cite{luethi2024fate}. The two topological phase transitions are determined by $\det \mathcal{H}(k=0) =0$ and $\det \mathcal{H}(k=\pi) = 0$, giving
\begin{subequations}  \label{eq:calculate_zeeman_phase_transition}
	\begin{align}
		0 &= ( \mu_{N,\mathrm{a}}^2 - \Delta_{Z,0}^2 )
		( \mu_{S,\mathrm{a}}^2 + \Delta^2 )
		\nonumber \\ &
		+ \! 8 t^2 \cos^2 \big(\frac{\Phi_\mathrm{SOI}}{2} \big)
		\big[
		2 t^2 \cos^2 \big(\frac{\Phi_\mathrm{SOI}}{2} \big)
		\! - \! \mu_{N,\mathrm{a}} \mu_{S,\mathrm{a}}
		\big]  , \\
		 0 &= ( \mu_{N,\mathrm{a}}^2 - \Delta_{Z,\pi}^2 )
		( \mu_{S,\mathrm{a}}^2 + \Delta^2 )
		\nonumber \\ & 
		+ \! 8 t^2 \sin^2 \big(\frac{\Phi_\mathrm{SOI}}{2} \big)
		\big[
		2 t^2 \sin^2 \big(\frac{\Phi_\mathrm{SOI}}{2}\big)
		\! -\!  \mu_{N,\mathrm{a}} \mu_{S,\mathrm{a}}
		\big]  ,
	\end{align}
\end{subequations}
where $\Delta_{Z,0}$ ($\Delta_{Z,\pi}$) indicates the Zeeman energy at which the bulk gap closes at $k=0$ ($k=\pi$). We label $\Delta_{Z,1} = \min \{\Delta_{Z,0}, \Delta_{Z,\pi}\}$ and $\Delta_{Z,2} = \max \{\Delta_{Z,0}, \Delta_{Z,\pi}\}$.
To determine whether the false PMMs is of the category ``before TP'' or ``after TP'', we compare the Zeeman energy at which the PMM is found to $\Delta_{Z,1/2}$. If $\Delta_Z < \Delta_{Z,1}$ ($\Delta_Z > \Delta_{Z,2}$), then the false PMM belongs to the category  ``before TP'' (``after TP'') and if $\Delta_{Z,1} \leq \Delta_Z \leq \Delta_{Z,2}$, then it is a true PMMs. 
This results in three distinct regions in the $t$ versus $\Phi_\mathrm{SOI}$ parameter space (with $\Delta_Z/\Delta$ of the PMMs fixed), see Figs.~\ref{fig:classification}(b) and (d). 
We find that the classification of numerically optimized PMMs with $\mu_{N,\mathrm{n}}$ and $\mu_{S,\mathrm{n}}$ agrees well with these three categories. 
As mentioned above, we do not find any false PMM of the category ``after TP, without ZEC''. This is also demonstrated in Fig.~\ref{fig:classification}. Such a PMM would require $\Delta_Z > \Delta_{Z,2}$ but we see from the figure that this condition also coincides with the region of parameter space where $|\Psi_{0,1\dn}|^2 \approx 1$, which, as explained above, results in PMMs of the category ``after TP, with ZEC''.

As explained above, if $|\Psi_{0,1\dn}|^2 \approx 1$ does not hold, then we do not necessarily expect the numerically determined chemical potentials $\mu_{N/S,\mathrm{n}}$ and the analytically calculated chemical potentials $\mu_{N/S,\mathrm{a}}$ to agree. As we show in Appendix~\ref{app:analytical_vs_optimized_mu}, the analytically calculated chemical potentials are still close to the numerically determined chemical potentials and therefore, the $\mu_{N/S,\mathrm{a}}$ found above analytically can still be used to classify the ``before TP'' versus ``after TP'' category in Fig. \ref{fig:classification}. From Fig.~\ref{fig:classification}, it also follows that PMMs are more numerous for $\mu_N > 0$ and $\mu_S > 0$ compared to $\mu_S<0$, and that, for $\mu_N, \mu_S>0$, the larger $t$ is, the more fine-tuning of $\Phi_\mathrm{SOI}$ is required to get PMMs. We will discuss the implications of this in Sec.~\ref{sec:prevalence_2_sites}.

\begin{figure}
	\centering
	\includegraphics[width=\linewidth]{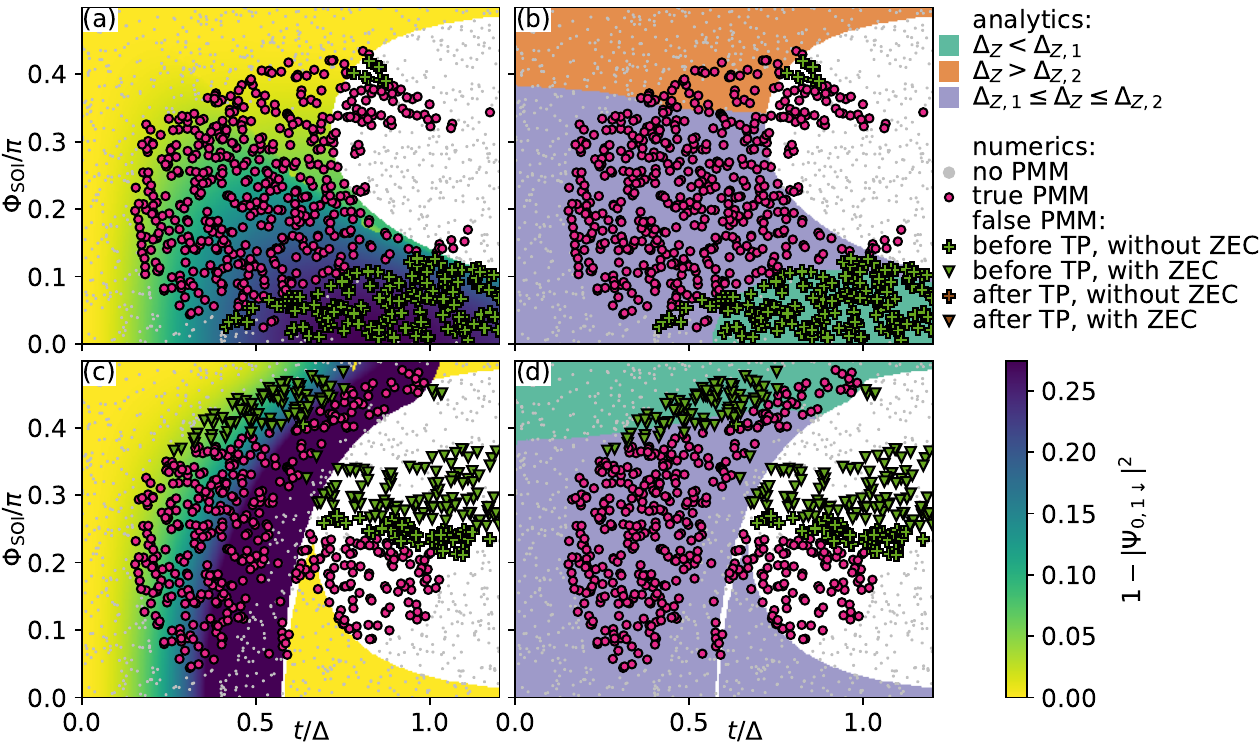}
	\caption{
	Classification of true (pink circles) and false (green and orange symbols) PMMs in a chain consisting of three normal and two superconducting QDs. This figure and its interpretation are equivalent to Fig.~\ref{fig:classification} but in a slightly longer chain. 
	In panel~(b), no false PMMs in the orange region with $\Delta_Z > \Delta_{Z,2}$ were found. Although there are some states in this region that have the long-chain behavior shown in Fig.~\ref{fig:false_pmms_examples}(e), these states do not satisfy the threshold conditions for PMMs in the three-site chain.
	In the white region, there is no analytical solution satisfying both Eqs.~\eqref{eq:mu_muM_eq1} and~\eqref{eq:mu_muM_3QDs}. 
	In contrast to the minimal chain, in the  three-site chain, there are also PMMs in these white regions, thus hinting that the physics of (false) PMMs can be richer in longer chains.
	The threshold values for the TR are $\Delta E_\mathrm{th}/\Delta = 10^{-3}$, $\Delta Q_\mathrm{th} = 0.1$, $M_\mathrm{th} = 0.3$, and $E_\mathrm{ex,th} = 0.02$.
	}
	\label{fig:classification_3QDs}
\end{figure}
\subsection{\label{sec:classification_three_site}False PMMs in three-site chains}
Next, we focus on a chain consisting of three normal QDs and two superconducting QDs, i.e., $N=3$ in Eq.~\eqref{eq:full_hamiltonian}. To solve for $\mu_{N,\mathrm{a}}$ and $\mu_{S,\mathrm{a}}$, the determinant of the BdG Hamiltonian defined in Eq.~\eqref{eq:full_hamiltonian} with $N=3$ is set to zero, thus Eq.~\eqref{eq:mu_muM_eq2} is replaced by
{\allowdisplaybreaks
\begin{align} \label{eq:mu_muM_3QDs}
0 =& 
( \Delta_Z^2 - \mu_{N,\mathrm{a}}^2 )^3  ( \Delta^2 + \mu_{S,\mathrm{a}}^2 )^2 
\nonumber \\ &
+ 8t^2 \mu_{N,\mathrm{a}} \mu_{S,\mathrm{a}} (\Delta_Z^2 - \mu_{N,\mathrm{a}}^2)^2 (\Delta^2 + \mu_{S,\mathrm{a}}^2)
\nonumber \\ &
+ t^4 \{-2 (\Delta_Z^2 - \mu_{N,\mathrm{a}}^2) [\Delta^2 (3\Delta_Z^2 - 5 \mu_{N,\mathrm{a}}^2)
\nonumber \\ &
+ \mu_{S,\mathrm{a}}^2 (3\Delta_Z^2 - 11 \mu_{N,\mathrm{a}}^2)] 
\nonumber \\ &
- 4 \Delta_Z^2 (\Delta_Z^2 - \mu_{N,\mathrm{a}}^2) (\Delta^2 + \mu_{S,\mathrm{a}}^2) \cos 2 \Phi_\mathrm{SOI}\}
\nonumber \\ &
+ t^6 [8 \mu_{N,\mathrm{a}} \mu_{S,\mathrm{a}} (-2\Delta_Z^2 + 3 \mu_{N,\mathrm{a}}^2)
\nonumber \\ &
 - 8 \Delta_Z^2 \mu_{N,\mathrm{a}} \mu_{S,\mathrm{a}} \cos 2 \Phi_\mathrm{SOI} ]
\nonumber \\ &
+ t^8 [3 (\Delta_Z^2 - 3 \mu_{N,\mathrm{a}}^2)
\nonumber \\ &
+ 4 \Delta_Z^2 \cos 2\Phi_\mathrm{SOI} + 2 \Delta_Z^2 \cos 4 \Phi_\mathrm{SOI} ] .
\end{align}
}
Solving Eqs.~\eqref{eq:mu_muM_eq1} and~\eqref{eq:mu_muM_3QDs} for $\mu_{N,\mathrm{a}}$ and $\mu_{S,\mathrm{a}}$ and following the same steps as above, we get a prediction on where in parameter space the different false PMMs categories are for $N=3$, see Fig.~\ref{fig:classification_3QDs}. We compare these results with numerically optimized PMMs with chemical potentials $\mu_{N,\mathrm{n}}$ and $\mu_{S,\mathrm{n}}$. Since the second-quantized Hamiltonian that is required for the optimization grows exponentially with the system size, doing the numerical optimization as explained in Ref.~\cite{luethi2024perfect} would be too slow here. Instead, we use SciPy's~\cite{2020SciPy-NMeth} ``minimize'' function to minimize the quantity $2 - M_L - M_R$ with the constraint $\Delta E = 0$. In contrast to the optimization algorithm of Ref.~\cite{luethi2024perfect}, the result might be a local, instead of a global, minimum. Nevertheless, this much faster algorithm is sufficient to get a general idea about the distribution and the properties of the PMMs in parameter space.

In general, the categories of the numerically optimized PMMs agree with the analytical predictions. Therefore, the analytical explanation for false PMMs, based on localized states in a double QD system without superconductivity, can also explain the occurrence of false PMMs in longer chains. However, the numerical and analytical results do not agree as well as was the case for the minimal chain. 
One reason for this might be the simplified optimization algorithm for $\mu_{N/S,\mathrm{n}}$ in the three-site chain, since it can get stuck in local minima.
Another explanation is that, as the chain is extended, there can be additional mechanisms that cause PMMs going beyond localized states in double QD systems. This claim is supported by the fact that true and false PMMs are found even in regions where $\mu_{N/S,\mathrm{a}}$ are not defined (white regions in Fig.~\ref{fig:classification_3QDs}). As discussed above, false PMMs without a ZEC do not have to solve Eq.~\eqref{eq:mu_muM_eq1} and thus it is not surprising that these states exist in regions of parameter space where Eqs.~\eqref{eq:mu_muM_eq1} and~\eqref{eq:mu_muM_3QDs} do not have a common solution. 
However, the fact that false PMMs with a ZEC exist in these regions of parameter space is an indication that the physics of false PMMs can be even richer in longer chains than what is considered in this section.

\section{\label{sec:prevalence_2_sites}Effect of threshold values on the ratio of true and false PMMs}
As was demonstrated in Ref.~\citenum{luethi2024fate}, it is difficult to distinguish false and true PMMs based solely on conductance measurements as they have similar properties. However, we found in Sec.~\ref{sec:classification} that true and false PMMs appear in relatively distinct regions of parameter space if $\Delta_Z/\Delta$ is fixed. Therefore, one might think that increased knowledge about the system parameters can help to distinguish true from false PMMs. 
However, the question arises whether distinct regions in the $t$ versus $\Phi_\mathrm{SOI}$ parameter space remain if $\Delta_Z$ is varied as well.
Furthermore, although well-defined in a theoretical model, $t$ and $\Phi_\mathrm{SOI}$ are not easily measurable parameters of experimental setups. 
Thus, in this section, we study whether true and false PMMs can be distinguished without having full knowledge of the parameter space and the effect of the threshold values on the ratio of true to false PMMs. In this section, we limit the discussion to minimal chains with $N=2$.

\begin{figure}
	\centering
	\includegraphics[width=\linewidth]{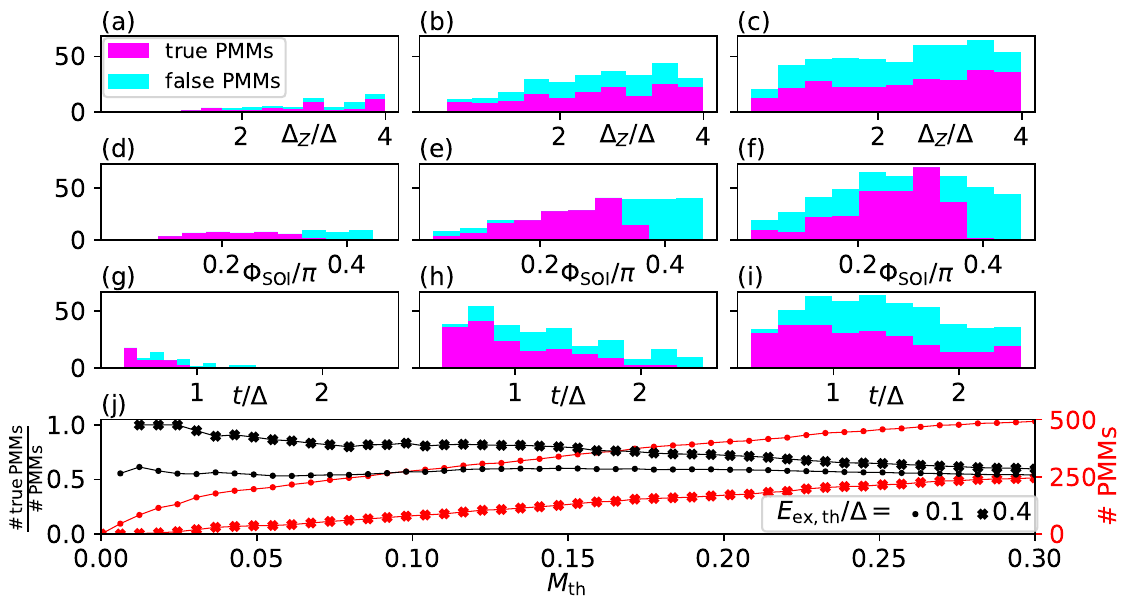}
	\caption{
		(a)-(i):~Histograms showing the Zeeman energies $\Delta_Z$ [panels (a)-(c)], SOI angles $\Phi_\mathrm{SOI}$ [panels (d)-(f)], and hopping amplitudes $t$ [panels (g)-(i)] of true and false PMMs in minimal chains. Note that the histograms of true and false PMMs are stacked.
		In panels (a), (d), and (g) the MP threshold is $M_\mathrm{th}=0.01$, in panels (b), (e), and (h) $M_\mathrm{th}=0.1$, and in panels (c), (f), and (i) $M_\mathrm{th}=0.3$.
		(j):~Ratio of true PMMs compared to the total number of PMMs, as a function of $M_\mathrm{th}$ (black symbols). The red symbols indicate the total number of PMMs. 
		The data indicated with dots (crosses) has a threshold value for the excitation gap of $E_\mathrm{ex,th}/\Delta=0.1$ ($E_\mathrm{ex,th}/\Delta=0.4$).
		The stricter the threshold values are, i.e., the smaller $M_\mathrm{th}$ or the larger $E_\mathrm{ex,th}$, the fewer states are overall classified as PMMs, but the more likely it is that these are true PMMs. 
		As the threshold values are relaxed, more states satisfy the TR condition defined in Eq.~\eqref{eq:definition_ROT} but the ratio of true PMMs decreases. 
		The threshold values are $\Delta E_\mathrm{th}/\Delta = 10^{-4}$, $\Delta Q_\mathrm{th}=0.01$, and $E_\mathrm{ex,th}/\Delta = 0.1$ for panels (a)-(i).
		The randomly selected values for the Zeeman energy are limited to $0 \leq \Delta_Z/\Delta \leq 4$, for the SOI angle $0 \leq \Phi_\mathrm{SOI} \leq \pi/2$, and for the hopping amplitude $0 \leq t/\Delta \leq 2.5$. 
		In total, 1500 distinct combinations of $(t, \Phi_\mathrm{SOI}, \Delta_Z)$ are used for this plot.
	}
	\label{fig:histogram_and_ratio_t_phiSOI_dZ}
\end{figure}

Whether a state is classified as a PMM depends on the threshold values chosen in Eq.~\eqref{eq:definition_ROT}. Therefore, it is natural to assume that the ratio of true to false PMMs also depends on the threshold values. 
To investigate this, we set random values for $\Delta_Z/\Delta$, $t/\Delta$, and $\Phi_\mathrm{SOI}$, then determine the chemical potentials $\mu_{N,\mathrm{n}}$ and $\mu_{S,\mathrm{n}}$ following the optimization algorithm described in Ref.~\citenum{luethi2024perfect} to search for PMMs. It is important to emphasize that this algorithm attempts to find the optimal chemical potentials such that the characteristics are closest to those of a perfect PMM, however, it does not rule out other (disconnected) regions of parameter space that could also satisfy the threshold conditions and result in an imperfect PMM. It should also be noted that, during the optimization process, we do not optimize for a large excitation gap $E_\mathrm{ex}$, but simply check after the optimization if the excitation gap exceeds the threshold value $E_\mathrm{ex,th}$.
As expected, the stricter the threshold values for the TR are, the fewer states are classified as PMMs. Neither $\Delta_Z$ nor $t$ alone are good indicators of whether a state is a true or false PMMs, see Fig.~\ref{fig:histogram_and_ratio_t_phiSOI_dZ}. In contrast, $\Phi_\mathrm{SOI}$, in combination with a low value for $M_\mathrm{th}$, seems to be useful to distinguish true from false PMMs, as PMMs that appear at low or high values of $\Phi_\mathrm{SOI}$ have a high chance of being false PMMs, whereas PMMs at intermediate values of $\Phi_\mathrm{SOI}$ are likely to be true PMMs.
This observation agrees well with the analytical understanding derived in Sec.~\ref{sec:classification}.

In certain cases, the ratio of true PMMs can reach 100\%, i.e., all states produced by the optimization algorithm that satisfy the TR condition are true PMMs, see Fig.~\ref{fig:histogram_and_ratio_t_phiSOI_dZ}(j). However, in these cases there are only very few states that satisfy the TR condition and therefore, the statistical significance of these results is low. In addition, more fine-tuning is required in this case to have a PMM in the system.
As the threshold values for the TR are relaxed, more states are classified as PMMs, but the ratio of true PMMs drops well below 100\%, see Fig.~\ref{fig:histogram_and_ratio_t_phiSOI_dZ}(j).

\begin{figure}
	\centering
	\includegraphics[width=\linewidth]{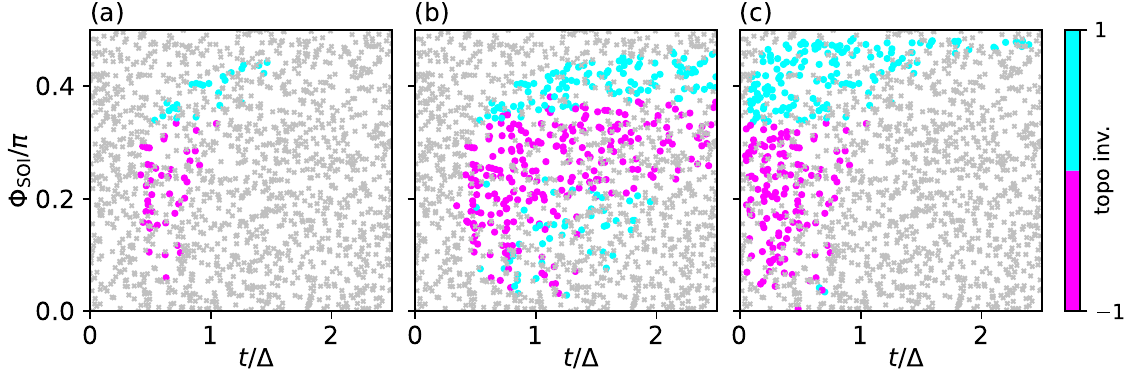}
	\caption{
		Scatter plot of PMMs in the $t$ versus $\Phi_\mathrm{SOI}$ parameter space, using the same data as for Fig.~\ref{fig:histogram_and_ratio_t_phiSOI_dZ}. The Zeeman energy $\Delta_Z$ of the states is not shown.
		Pink (blue) points indicate true (false) PMMs, and silver crosses indicates states that do not satisfy the TR condition defined in Eq.~\eqref{eq:definition_ROT} after the optimization. 
		The threshold values are $\Delta E_\mathrm{th}/\Delta = 10^{-4}$, $\Delta Q_\mathrm{th} = 0.01$ in all panels, whereas  (a) $M_\mathrm{th}=0.01$ and $E_\mathrm{ex,th}/\Delta=0.1$, (b) $M_\mathrm{th}=0.2$ and $E_\mathrm{ex,th}/\Delta=0.1$, (c) $M_\mathrm{th}=0.01$ and $E_\mathrm{ex,th}/\Delta=10^{-3}$.
		Zero-energy states at small hopping amplitudes $t$ have a small excitation gap and therefore, these states satisfy the TR condition only if $E_\mathrm{ex, th}$ is small enough. Zero-energy states at large $t$ have a higher MP and therefore they only satisfy the TR condition if $M_\mathrm{th}$ is large enough. 
	}
	\label{fig:scatter_no_U}
\end{figure}

\begin{figure}
	\centering
	\includegraphics[width=\linewidth]{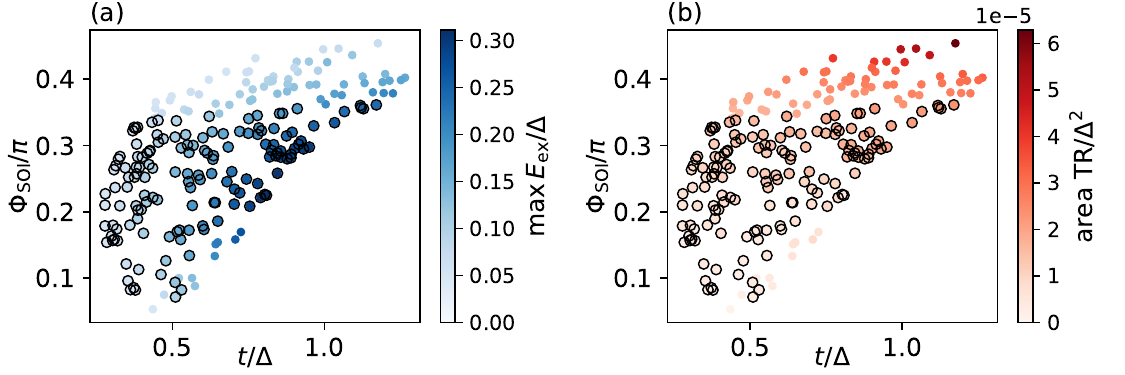}
	\caption{
	(a): Largest excitation gap $E_\mathrm{ex}$ in the TR and (b): area of the TR in the $\mu_N$ versus $\mu_S$ parameter space for PMMs that are found using the numerical optimization algorithm described in Ref.~\cite{luethi2024perfect} with $\Delta_Z/\Delta=0.8$ fixed.
	In both panels, the points circled in black are the true PMMs, the one without a black circle are the false PMMs. 
	Although the area of the TR is highly dependent on the threshold values, comparing its size at fixed threshold values is an indication of the stability of the PMM, on the feasibility of braiding, and on the required amount of tine-tuning of the chemical potentials. Ideally, a PMM has a large excitation gap and large area of the TR, which
 is the case for large hopping amplitudes $t$. 
	The TR threshold values here are $\Delta E_\mathrm{th}/\Delta=10^{-4}$, $M_\mathrm{th}=0.1$, $\Delta Q_\mathrm{th}=0.01$, and $E_\mathrm{ex,th}/\Delta=0.05$.
	}
	\label{fig:gap_and_ROT_area}
\end{figure}

To better understand how the classification of PMMs depends on the threshold values, we study the distribution of true and false PMMs in the $t$ versus $\Phi_\mathrm{SOI}$ parameter space for different threshold values, see Fig.~\ref{fig:scatter_no_U}. 
States at small $t$ have a small excitation gap $E_\mathrm{ex}$, see Fig.~\ref{fig:gap_and_ROT_area}(a), therefore they are not classified as PMMs if $E_\mathrm{ex,th}$ is set too high. 
As $M_\mathrm{th}$ increases, states at larger $t$ satisfy the TR condition. These states at higher $t$ tend to have larger excitation gaps $E_\mathrm{ex}$. In addition, we find that the area of the TR in the $\mu_N$ versus $\mu_S$ parameter space tends to be larger for larger $t$, see Fig.~\ref{fig:gap_and_ROT_area}(b). A larger TR in the $\mu_N$ versus $\mu_S$ parameter space means that less fine-tuning of the chemical potential is required and it is easier to manipulate the states when doing, e.g., braiding~\cite{boross2023braiding, tsintzis2023roadmap, geier2023fermionparity}.
Therefore, true PMMs in systems with large hopping amplitudes $t$ seem ideal candidates for stable MBS-like states. However, as we have discussed in Sec~\ref{sec:classification}, the larger $t$, the more one has to fine-tune the SOI angle $\Phi_\mathrm{SOI}$ to enable PMMs, which becomes even more complicated if there is some interdependence between parameters. Furthermore, states at large $t$ generally have a lower MP and therefore they are only classified as PMMs if the threshold values are relaxed sufficiently, which leads to a lower ratio of true PMMs, see Fig.~\ref{fig:histogram_and_ratio_t_phiSOI_dZ}(j).
As such, the two main concerns for PMMs -- stability and a low chance for false PMMs -- are contradictory and one cannot guarantee that stable PMMs are connected to topological states in the long chain limit.

\section{\label{sec:conclusion}Conclusion}
We have studied false PMMs in chains consisting of alternating normal and superconducting QDs. In particular, based on a system without any superconductivity, we explained why false PMMs with certain characteristics appear in different regions of parameter space and we have found good agreement between this analytic understanding and numerical results. This emphasizes that false PMMs can occur in short chains, without any relation to MBSs that appear due to topological superconductivity in long chains.

Additionally, we demonstrated that the SOI angle $\Phi_\mathrm{SOI}$ has the potential to distinguish regions of parameter space where the number of true false PMMs is significantly larger than false PMMs and that the ratio of true to false PMMs depends on the threshold values that are used to define PMMs. The stricter the threshold values are set, the more probable it is that a state that fulfills the TR conditions is a true PMMs. However, the stricter the threshold values are, the fewer states overall are classified as PMMs, thus requiring more fine-tuning of parameters to reach the TR.

We have demonstrated that the most stable PMMs occur at large hopping amplitudes, as these states tend to have larger excitation gaps, require less fine-tuning of the chemical potential, and allow for more flexibility during braiding. However, states at large hopping amplitudes require more fine-tuning of the SOI angle and the localization is generally worse than for states at smaller hopping amplitudes. Therefore, the threshold values must be set rather loose to allow for stable PMMs.
However, we have also shown that the ratio of false PMMs increases as the threshold values are relaxed.
If instead, one were to set stricter threshold values, then the ratio of true PMMs would increase, but we find that such states only occur at smaller hopping amplitudes, where the excitation gaps tend to be smaller and more fine-tuning of the chemical potential is required to stay in the TR. Therefore, the regions of parameter space where stable PMMs occur and the regions with a low chance of false PMMs do not have a significant overlap, requiring a tradeoff.
Furthermore, we have so far assumed that the hopping amplitude $t$ and the SOI angle $\Phi_\mathrm{SOI}$ are independent parameters of our model. In reality, however, it is to be expected that there is some dependency between these parameters, thus making it even harder to tune the system into a region of parameter space hosting true PMMs.

\begin{acknowledgments}
This work was supported as a part of NCCR SPIN, a National Centre of Competence in Research, funded by the Swiss National Science Foundation (grant number 225153).
This project received funding from the European Union’s Horizon 2020 research and innovation program (ERC Starting Grant, Grant Agreement No. 757725).

\end{acknowledgments}

\section*{Data availability}
The data and code underpinning the results of this work is available in the following Zenodo repository: \href{https://doi.org/10.5281/zenodo.15180571}{10.5281/zenodo.15180571}.

\appendix

\section{\label{app:large_zeeman}Zeeman energy on the middle QD and larger Zeeman energy}
\begin{figure}
	\centering
	\includegraphics[width=\linewidth]{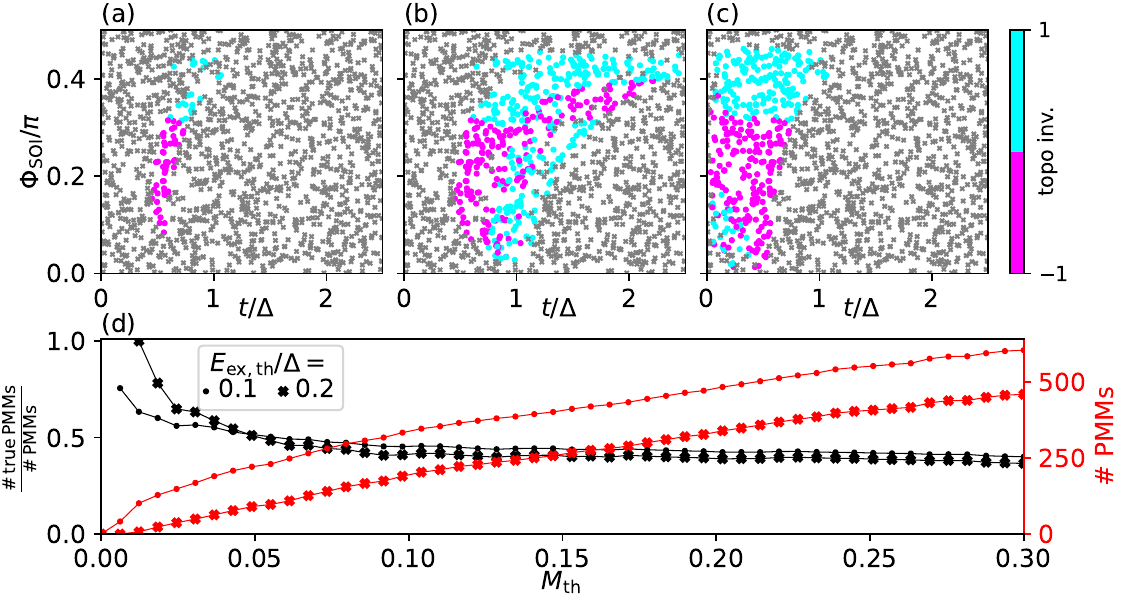}
	\caption{
		Classification of true and false PMMs in a minimal chain with $\Delta_Z/\Delta = 2$ and $\Delta_{Z,S}/\Delta=0.4$, see Eq.~\eqref{eq:definition_zeeman_M}. This figure and its interpretation are equivalent to Figs.~\ref{fig:histogram_and_ratio_t_phiSOI_dZ}(j) and~\ref{fig:scatter_no_U}, but with different Zeeman energies. The larger Zeeman energy on the normal QDs and non-zero Zeeman energy on the superconducting QDs do not lead to any qualitative difference.
		The threshold values for panels (a)-(c) are the same as for Figs.~\ref{fig:scatter_no_U}(a)-(c), respectively, and the threshold values for panel (d) are the same as for Fig.~\ref{fig:histogram_and_ratio_t_phiSOI_dZ}(j), except for $E_\mathrm{ex,th}$, whose values are given in the legend.
	}
	\label{fig:scatter_large_zeeman}
\end{figure}

In this section we adapt Eq.~\eqref{eq:definition_delta_z} as follows:
\begin{equation} \label{eq:definition_zeeman_M}
\Delta_{Z,j} = \begin{cases}
\Delta_Z & \mathrm{if} \, j \, \mathrm{even} , \\
\Delta_{Z,S} & \mathrm{if} \, j \, \mathrm{odd}  ,
\end{cases}
\end{equation}
i.e., we introduce a non-zero Zeeman energy on the superconducting QDs. Furthermore, we increase the Zeeman energy on the normal QDs. We consider the minimal chain ($N=2$) and find that false PMMs exist in systems with $\Delta_{Z,S}/\Delta=0.4$ and $\Delta_Z/\Delta=2.0$, see Fig.~\ref{fig:scatter_large_zeeman}. The ratio of true to false PMMs behaves qualitatively the same as for the data shown in the main text, where $\Delta_{Z,S}=0$ and $\Delta_Z/\Delta=0.8$.

\section{\label{app:with_U}Finite on-site Coulomb interaction}
In the main text, we set $U=0$ and could consequently use the BdG formulation of the Hamiltonian defined in Eq.~\eqref{eq:full_hamiltonian} for the long and infinite chain limits~\cite{luethi2024fate}. In this formulation, the topological invariant is defined as a Pfaffian.

In Ref.~\cite{kitaev2001unpaired}, the Majorana number $\mathcal{M}(H) = \pm 1$ was introduces. Having $\mathcal{M}(H) = -1$ ($+1$) means that the system is in the topological (trivial) phase and it is defined as
\begin{equation} \label{eq:majorana_number}
	P[
	H(L_1 + L_2)
	]
	= \mathcal{M} (H)
	P [
	H (L_1)
	]
	P [
	H (L_2)
	],
\end{equation}
where $P[H(L)]$ is the particle number parity of the ground state of the Hamiltonian $H(L)$ that describes a system of length $L$ with periodic boundary conditions. It is shown in Ref.~\cite{kitaev2001unpaired} that in noninteracting systems, the Majorana number is equivalent to the topological invariant derived from the Pfaffian.
However, if $U > 0$, the system is interacting, and therefore, the Pfaffian can no longer be used to calculate the topological invariant. Nevertheless, Eq.~\eqref{eq:majorana_number} remains valid~\cite{kitaev2001unpaired}.
We set $L_1=L_2 \equiv L$, thus Eq.~\eqref{eq:majorana_number} becomes $P[H(2 L)]= \mathcal{M} (H)P [H (L)]^2 =  \mathcal{M} (H)$. 
However, the length $L$ should be big compared to the MBS localization length, thus requiring DMRG~\cite{steven1992density, steven1993density, schollwock2005density} for systems with $U>0$. This is problematic since DMRG fares badly with periodic boundary conditions~\cite{steven1993density}, taking much longer to converge.
Thus, even with DMRG, we keep $L$ relatively short, which may lead to misclassifications. To double-check we also calculate $\Delta E (\Delta_Z)$ in a longer chain using DMRG in a system with open boundary conditions, as was done in Ref.~\cite{luethi2024fate}.
We emphasize that this does not guarantee the absence of misclassifications, but we assume that there is not a statistically significant number of them.

\begin{figure}
	\centering
	\includegraphics[width=\linewidth]{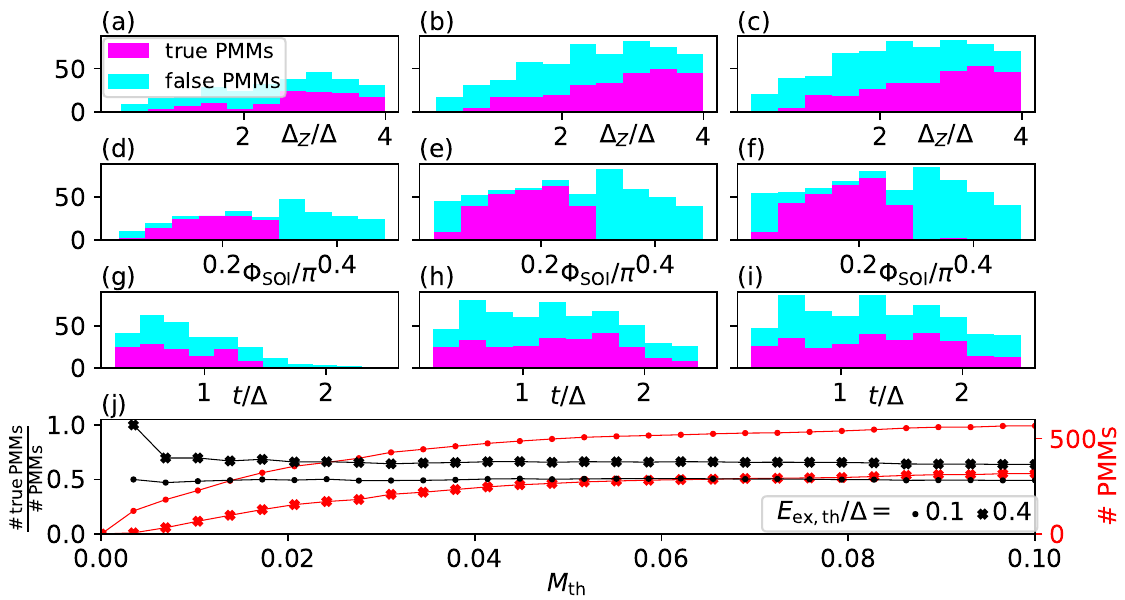}
	\caption{
		Histograms and ratio of true and false PMMs in a minimal chain, equivalent to Fig.~\ref{fig:histogram_and_ratio_t_phiSOI_dZ}, except for the on-site Coulomb repulsion which is here $U/\Delta=100$.
		There is no qualitative difference to the case $U=0$.
		The threshold values are $\Delta E_\mathrm{th}/\Delta = 10^{-4}$, $\Delta Q_\mathrm{th} = 0.01$, $E_\mathrm{ex,th}/\Delta=0.05$, $M_\mathrm{th}=0.01$ in panels (a), (d), (g), $M_\mathrm{th}=0.05$ in panels (b), (e), (h), and $M_\mathrm{th}=0.1$ in panels (c), (f), (l).
		In total, 1401 distinct combinations of $(t, \Phi_\mathrm{SOI}, \Delta_Z)$ were used for this plot.
	}
	\label{fig:histogram_and_ratio_t_phiSOI_dZ_with_U}
\end{figure}

In this Appendix, we set $U/\Delta = 100$, then continue as in the main text, i.e., we randomly choose a point in the $(\Delta_Z/\Delta, t/\Delta, \Phi_\mathrm{SOI})$ parameter space and use the optimization algorithm described in Appendix~D of Ref.~\cite{luethi2024perfect} to tune $\mu_N$ and $\mu_S$. We then check if the state satisfies the TR condition given in Eq.~\eqref{eq:definition_ROT}. If they do, then we calculate the Majorana number $\mathcal{M}$ according to Eq.~\eqref{eq:majorana_number} to determine whether the state is a true or false PMMs. The conclusions are qualitatively the same as for $U=0$ in the main text, see Fig.~\ref{fig:histogram_and_ratio_t_phiSOI_dZ_with_U}.

\section{\label{app:analytical_vs_optimized_mu}Comparing the analytically calculated chemical potentials with the numerically optimized chemical potentials}
\begin{figure}
	\centering
	\includegraphics[width=\linewidth]{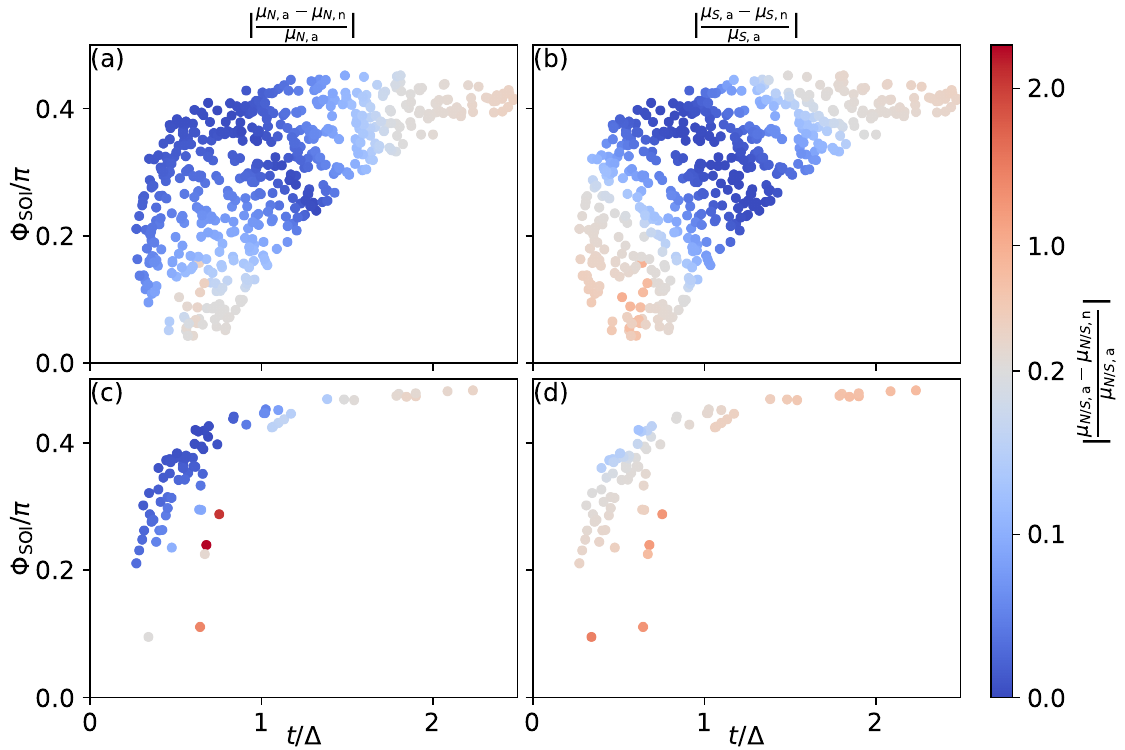}
	\caption{
		Relative difference between the analytically calculated chemical potentials $\mu_{N/S,\mathrm{a}}$ and the numerically determined chemical potentials $\mu_{N/S,\mathrm{n}}$ (see Sec.~\ref{sec:classification} for more detail).
		(a) and (b) [(c) and (d)]: Difference of chemical potentials for all PMMs of Fig.~\ref{fig:classification}(a) and (b) [(c) and (d)].
		(a) and (c) [(b) and (d)]: Difference between the chemical potentials on the normal [superconducting] QDs.
		Note the non-linear scale of the color map.
	}
	\label{fig:difference_chemical_potential}
\end{figure}

In this section, we compare the chemical potentials $\mu_{N,\mathrm{n}}$ and $\mu_{S,\mathrm{n}}$ determined via the optimization algorithm explained in Ref.~\cite{luethi2024perfect} with the analytically calculated chemical potentials $\mu_{N,\mathrm{a}}$ and $\mu_{S,\mathrm{a}}$ that are calculated as explained in Sec.~\ref{sec:classification}. We show this difference in Fig.~\ref{fig:difference_chemical_potential} for all states classified as PMMs in Fig.~\ref{fig:classification} (in the minimal chain).
In most cases, the relative differences between the analytic and numeric chemical potentials is less than 20\%. However, for small $\Phi_\mathrm{SOI}$, the difference increases. As explained in the main text, in this region of parameter space, the PMMs are no longer associated with a ZEC and therefore, the analytical constraints on the chemical potentials derived in Sec.~\ref{sec:classification} no longer necessarily hold. Furthermore, as $\Phi_\mathrm{SOI} \rightarrow \frac{\pi}{2}$ and $t/\Delta > 1$, the difference also grows. The reason for this is that the analytically calculated chemical potentials diverge in this region of parameter space, whereas the numerically optimized chemical potentials are bounded due to practical reasons.

\FloatBarrier

\bibliography{bibliography}

\end{document}